\documentclass[prl,twocolumn,superscriptaddress,letterpaper,amssymb,amsmath,floatfix]{revtex4-1}
\pdfoutput=1
\usepackage{amsmath}
\usepackage{stmaryrd}
\usepackage{color}
\usepackage{graphicx}
\usepackage{mathrsfs}
\usepackage{ textcomp }
\begin{document}

\title{Bilayer Twisted Hubbard Model for Sr$_3$Ir$_2$O$_7$}

\author{Jia-Wei Mei}
\affiliation{Institute for Theoretical Physics, ETH Z\"urich, 8093 Z\"urich, Switzerland}

\date{\today}

\begin{abstract}
  We use the bilayer twisted Hubbard model to study the low energy electronic structure of Sr$_3$Ir$_2$O$_7$. Sr$_3$Ir$_2$O$_7$ is suggested as a Mott insulator and described by the bilayer twisted pseudospin-1/2 model. The bilayer twists in Sr$_3$Ir$_2$O$_7$ bring the intrinsic anisotropy and the system is no longer SU(2) invariant. The anisotropy selects an easy $c$-axis collinear antiferromagnetic magnetic structure at low temperatures which breaks the discrete Ising symmetry. The transverse spin excitations are gapped and the longitudinal one is weakly dispersive. We implement the bond-operator mean field calculation to describe the magnetic properties of the bilayer twisted pseudospin-1/2 model in good agreement with experimental measurements in Sr$_3$Ir$_2$O$_7$. The bilayer twisted Hubbard model is believed to be the good zeroth approximation for the low energy electronic structure of Sr$_3$Ir$_2$O$_7$. 
\end{abstract}

\maketitle
\textit{Introduction.}
The strong spin-orbit coupling in Ir$^{4+}$ ($\sim$ 0.4 eV) yields a narrow half-filled $J_{\text{eff}}=1/2$ single band near the Fermi level in Ir-oxides and enhances the correlation effects of the one-site Coulomb repulsion ($\sim$ 2 eV)\cite{Kim2008,Jin2009,Clancy2012,Watanabe2010}. It brings  strongly correlated physics and makes Ir-oxides behave like Mott insulators\cite{Kim2008,Moon2008,Kim2009,Okabe2011,Okabe2011a,Liu2011,Singh2012}. The perovskite iridates are isostructural to  the cuprate parent compounds and have been drawing increasing attention in recent years. Compared with the cuprates, the iridates are in the vicinity of the Mott transition\cite{Moon2008,Okabe2011,Haskel2012,Arita2012}. With the help of the rapid-developing resonant x-ray magnetic scattering technique\cite{Ament2011}, the low energy properties of the iridates are well studied in the experiments\cite{Kim2008,Clancy2012,Moon2008,Kim2009,Kim2012a,Kim2012b,Boseggia2012,Kim2012b,Clancy2012a,Fujiyama2012}. The iridates provide the platform for investigation of the low energy electronic structures of the Mott physics. The similarity and distinction between the cuprates and perovskite iridates can improve our understanding of the Mott physics on the two dimensional square lattice.

\begin{figure}[b]
  \begin{center}
    \includegraphics[width=0.6\columnwidth]{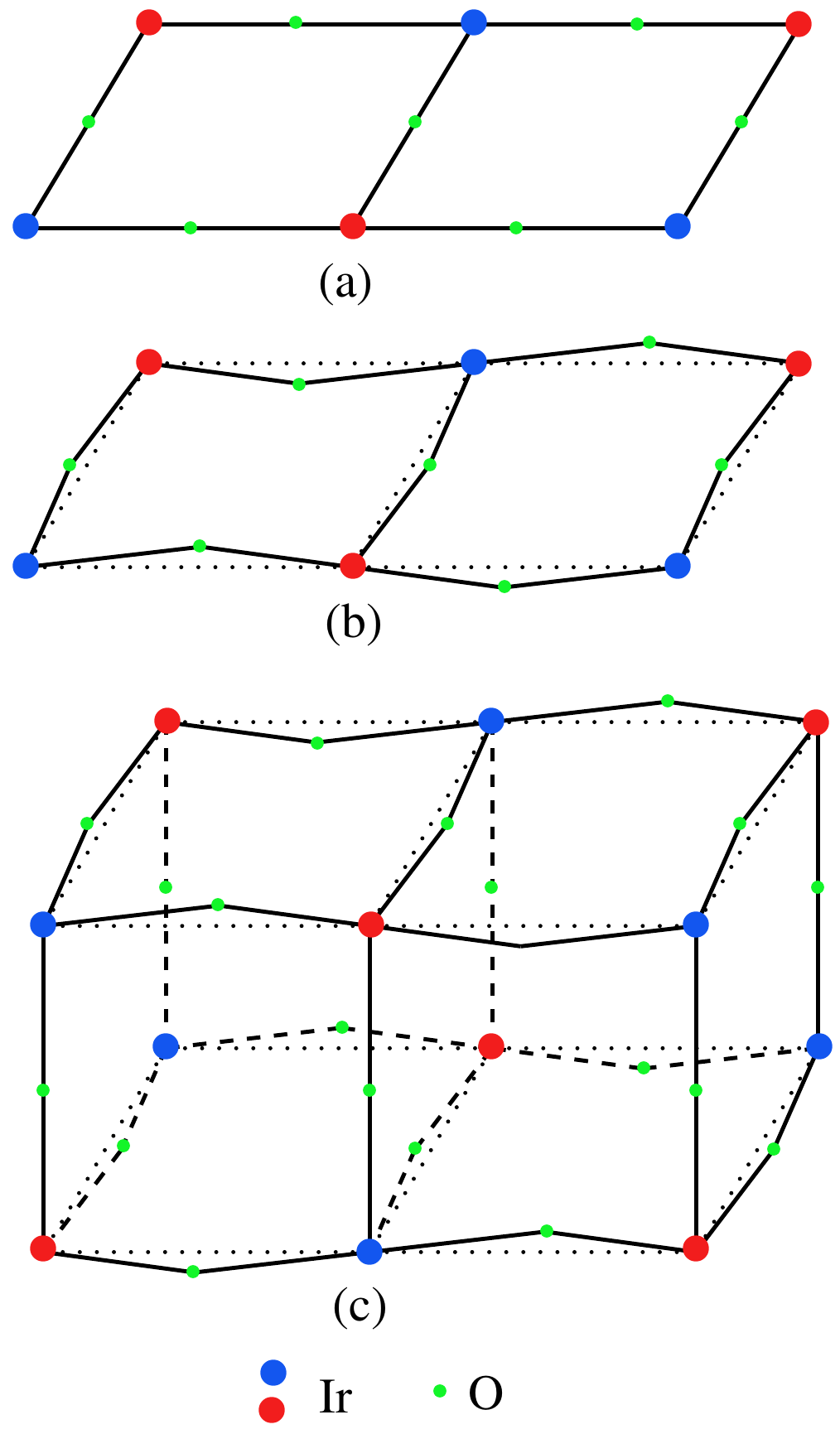}
  \end{center}
  \caption{Schematic crystal structures of IrO$_2$ planes in (a) Ba$_2$IrO$_4$, (b) Sr$_2$IrO$_4$ and (c)  Sr$_3$Ir$_2$O$_7$.}
  \label{fig:Sr327}
\end{figure}

The perovskite iridates have IrO$_2$ planes which determine the low energy electronic properties, similar to the parent compound of cuprates. Ba$_2$IrO$_4$ and Sr$_2$IrO$_4$ have the single layer of IrO$_2$ plane. As shown in Fig. \ref{fig:Sr327} (a) and (b), the Ir-O-Ir bonds in the IrO$_2$ plane are straight for Ba$_2$IrO$_4$ and alternating twisted for Sr$_2$IO$_4$, respectively. The low energy magnetic properties of Ba$_2$IrO$_4$ are the same as those of La$_2$CuO$_4$\cite{Okabe2011,Okabe2011a}. Sr$_2$IrO$_4$ has the in-plane canted antiferromagnetic structure and shows ferromagnetism with large ferromagnetic moment due to the twists of Ir-O-Ir bonds\cite{Cao1998,Wang2011}. The twists in Sr$_2$IrO$_4$ can be gauged away and the magnetic excitations are similar to La$_2$CuO$_4$\cite{Wang2011,Jackeli2009,Kim2012}.  The activation energy gaps in electrical resistivity are estimated to be $\sim$ 0.07 eV for Ba$_2$IrO$_4$\cite{Okabe2011} and $\sim$ 0.06 eV for Sr$_2$IrO$_4$\cite{Shimura1995}, respectively.  The perovskite Ir-oxides are in the vicinity of the Mott transition and the pressure can drive the systems cross the transition\cite{Okabe2011,Haskel2012,Arita2012}.   

Sr$_3$Ir$_2$O$_7$ is the bilayer perovskite iridate with  strongly coupled bilayers of IrO$_2$ planes, stacked along the $c$-axis with pairs of adjacent bilayers  offset by half a unit cell\cite{Cao2002}.  As shown in Fig. \ref{fig:Sr327} (c), the Ir-O-Ir bonds in the bilayer IrO$_2$ planes are twisted. The twisted directions are alternating within each layer and also opposite between the two layers\cite{Cao2002,Fujiyama2012}. The optical gap in Sr$_3$Ir$_2$O$_7$ is finite although small\cite{Moon2008}, indicating the Mottness.  Recently several measurements have been performed to explore the low temperature magnetic properties of Sr$_3$Ir$_2$O$_7$\cite{Boseggia2012,Kim2012a,Kim2012b,Clancy2012a,Fujiyama2012,Dhital2012}.  At low temperatures, the system has the long-range antiferromagnetic (AF) order as probed in the x-ray  magnetic Bragg scattering\cite{Boseggia2012,Kim2012a,Kim2012b,Clancy2012a,Fujiyama2012} and the neutron scattering\cite{Dhital2012}. The magnetic order has the easy $c$-axis collinear AF structure\cite{Kim2012a,Fujiyama2012}, in the contrast to the single layer Sr$_2$IrO$_4$ with in-plane canted antiferromagnetic structure\cite{Cao1998}.  The transverse spin wave excitations are gapped in Sr$_3$Ir$_2$O$_7$\cite{Kim2012b}, in the contrast to the gapless  Sr$_2$IrO$_4$\cite{Kim2012}. There is a weak and broad feature above the transverse band in the inelastic x-ray scattering measurement in Sr$_3$Ir$_2$O$_7$\cite{Kim2012b}. The magnetic moment in Sr$_3$Ir$_2$O$_7$ is weaker than that in Sr$_2$IrO$_4$\cite{Fujiyama2012}. Increasing the temperature, the Bragg peaks of the AF order disappear at the magnetic transition temperature, $T_N\sim 280$ K\cite{Fujiyama2012,Dhital2012}.  

In this paper, we will use the bilayer twisted Hubbard model to describe the low energy properties of Sr$_3$Ir$_2$O$_7$. Due to the weak antiferromagnetism in Sr$_3$Ir$_2$O$_7$, the Mottness is questioned and the itinerant band magnet is suggested\cite{Carter2012,Fujiyama2012}. However, we insist on  the Mott scenario for Sr$_3$Ir$_2$O$_7$. Although small, the optical gap in Sr$_3$Ir$_2$O$_7$ is definitely finite\cite{Moon2008}. The large spin gap is not compatible  with the weak magnetic moment in the itinerant band magnet\cite{Kim2012b}. Even if Sr$_3$Ir$_2$O$_7$ is the spin-density-wave band magnet, the magnetism has the same symmetry as that raised from the local pseudospins and the Mott scenario is still enlightening for the understanding of the magnetic structures. Due to the Mottness, we study the bilayer twisted pseudospin-1/2 model. The bilayer twists on Ir-O-Ir bonds in Sr$_3$Ir$_2$O$_7$ can not be gauged away and bring the intrinsic anisotropy. The anisotropy selects the easy $c$-axis antiferromagnetic structure at low temperatures in which the discrete Ising symmetry is broken.  We implement the bond-operator mean field calculation to quantitatively describe the magnetic properties. The transverse spin wave excitations are gapped and the longitudinal one is weakly dispersive. We interpret the weak and broad feature above the transverse band as the weakly dispersive longitudinal spin wave excitations rather than the two-magnon scattering\cite{Kim2012b}.  The interlayer coupling increases the itinerant of the electron and reduces the AF moment in Sr$_3$Ir$_2$O$_7$. On the mean field level, we ignore the quantum fluctuations and the spin gap and the magnetic moment reduction are significantly underestimated. For all that, however, the bilayer twisted Hubbard model still gives a good qualitative description of the magnetic structures of Sr$_3$Ir$_2$O$_7$.

\textit{Bilayer twisted Hubbard model for Sr$_3$Ir$_2$O$_7$.}
The crystal field in  IrO$_6$ octahedra  splits the $d$-orbitals of Ir$^{4+}$ into  $e_g$ levels and $t_{2g}$ levels. The active $t_{2g}$ levels in Ir$^{4+}$ has an effective angular momentum $l=1$, $|l_z=0\rangle\equiv|XY\rangle$, $|l_z=\pm1\rangle\equiv-\frac{1}{\sqrt{2}}(i|XZ\rangle\pm|YZ\rangle)$ and is split further into $J_{\text{eff}}=3/2$ and $J_{\text{eff}}=1/2$ by the on-site spin-orbit coupling.  The ratios of the out-of-plane Ir-O and in-plane Ir-O bond lengths are 1.07 for Ba$_2$IrO$_4$, 1.04 for Sr$_2$IrO$_4$ and 1.02 for Sr$_3$Ir$_2$O$_7$\cite{Okabe2011,Cao1998,Cao2002}. The elongation of the IrO$_6$ along $c$-axis is negligible and Ir$^{4+}$ has the ideal $J_{\text{eff}}=1/2$ states as
\begin{eqnarray}
  |J_{\text{eff}}^z&=& +1/2\rangle=\frac{1}{\sqrt{3}}(|0,\uparrow\rangle-\sqrt{2}|+1,\downarrow\rangle),\nonumber\\
  |J_{\text{eff}}^z&=& -1/2\rangle=\frac{1}{\sqrt{3}}(|0,\downarrow\rangle-\sqrt{2}|-1,\uparrow\rangle).
\end{eqnarray}
As emphasized in Ref. \onlinecite{Wang2011}, $|XY\rangle$, $|YZ\rangle$ and $|XZ\rangle$ are the $t_{2g}$ orbitals defined in the local cubic axis of the IrO$_6$ octahera. 

In the IrO$_2$ planes, the hoppings of $J_{\text{eff}}=1/2$ electrons between Ir sites are mediated by the oxygen $2p$ orbitals.  For Ba$_2$IrO$_4$ as shown in Fig. \ref{fig:Sr327} (a), the Ir-O-Ir bonds are straight and the hopping term for  $J_{\text{eff}}=1/2$ electron on the Ir-O-Ir bond $\mathbf{r}_{ij}$ is given as
\[h_{ij}=-t\sum_{\sigma} (c_{i\sigma}^\dag c_{j\sigma}+\text{h.c.}),\]
where $c_{i\sigma}$ is the electron operator on the $i$-th site with $\sigma=\uparrow/\downarrow$ for $J_{\text{eff}}^z=\pm1/2$. For the straight Ir-O-Ir bonds the local cubic axis are also the global ones and the hopping parameter $t$ can be specified as the real number. 

For Sr$_2$IrO$_4$ as shown in Fig. \ref{fig:Sr327} (b), the Ir-O-Ir bonds are twisted and the local cubic axis are alternatively rotated respect to the global axis. The twisted angle is $\theta\sim11$\textdegree.  Here we still interpret $c_{i\sigma}$  as the electron operator based on the local cubic axis\footnote{It should be noted that the electron operator is based on the local cubic axis in Ref. \onlinecite{Jin2009} and on the global cubic axis in Ref. \onlinecite{Wang2011}, respectively.}. On the twisted Ir-O-Ir bonds, the hopping term is now\cite{Jin2009,Wang2011,Note1}
\begin{eqnarray}
  h_{ij}^{\text{twist}}=-t\sum_{\sigma}(e^{-i\epsilon_i\epsilon_\sigma\theta}c_{i\sigma}^\dag c_{j\sigma}+\text{h.c.}),
\end{eqnarray}
with $\epsilon_i=\pm1$ for the sublattice and $\epsilon_\sigma=\pm1$ for $\sigma=\uparrow/\downarrow$. The hopping parameter is the complex number, $t e^{-i\epsilon_i\epsilon_\sigma\theta}$. The twists of the single layer Sr$_2$IrO$_4$ can be removed by proper site-dependent spin rotations and the twisted Hubbard model can be mapped to the SU(2)-invariant pseudospin-1/2 model\cite{Wang2011,Jackeli2009} which has the magnetic spectrum mimic that in the cuprate\cite{Kim2012}.

For Sr$_3$Ir$_2$O$_7$ as shown in Fig. \ref{fig:Sr327} (c), we can read off the bilayer twisted Hubbard model similar to the above construction
\begin{eqnarray}
  H=\sum_{\langle ij\rangle l}h_{ij}^l+\sum_{i}h_i^\bot+U\sum_{li}c_{li\uparrow}^\dag c_{li\uparrow} c_{li\downarrow}^\dag c_{li\downarrow},
\end{eqnarray}
where $\langle ij\rangle$ denotes  the nearest neighbor bonds in $l$-th ($l=1,2$) layer. The hopping on the twisted bonds for each layer is
\[h_{ij}^l= -t\sum_{\sigma}(e^{-i\epsilon_l\epsilon_i\epsilon_\sigma\theta}c_{li\sigma}^\dag c_{lj\sigma}+\text{h.c.}),\]
with $\epsilon_l=\pm1$ for the layer. The interlayer Ir-O-Ir bonds are straight in Sr$_3$Ir$_2$O$_7$ and the hopping on the interlayer rungs along $c$-axis is
\[h_{i}^\bot= -t_{\bot}\sum_{\sigma}(c_{1i\sigma}^\dag c_{2i\sigma}+\text{h.c.}).\]
For Sr$_3$Ir$_2$O$_7$, the twist angle is $\theta\simeq12$\textdegree. The hopping parameters ($t$ and $t_\bot$) are around 0.2 eV and the on-site repulsive interaction $U$ takes the value around 2 eV analogous to those in Sr$_2$IrO$_4$\cite{Jin2009}.

Based on the experiments (finite optical gap and large spin gap), we use the Mott scenario and obtain the bilayer twisted pseudospin-1/2 model for Sr$_3$Ir$_2$O$_7$
\begin{eqnarray}
  \label{eq:bth}
  H&=& J\sum_{\langle ij\rangle l}[\cos(2\theta)\mathbf{S}_{li}\cdot\mathbf{S}_{lj}+2\sin^2(\theta)S_{li}^zS_{lj}^z\nonumber\\
    &&-\epsilon_i\epsilon_{l}\sin(2\theta)(\mathbf{S}_{li}\times\mathbf{S}_{lj})\cdot\hat{z}]+J_{\bot}\sum_{i}\mathbf{S}_{1i}\cdot\mathbf{S}_{2i}.
\end{eqnarray}
The super-exchange constants are given as $J=4t^2/U$ and $J_{\bot}=4t_{\bot}^2/U$, $J\simeq J_\bot\simeq80$ meV for Sr$_3$Ir$_2$O$_7$. We can implement the site-dependent rotations and interpret the pseudospin in the global cubic axis
\begin{eqnarray}
  \begin{pmatrix}\tilde{S}_{li}^x\\\tilde{S}_{li}^y\\\tilde{S}_{li}^z\end{pmatrix}=\begin{pmatrix}\cos(\theta)&\epsilon_{l}\epsilon_i\sin(\theta)&0\\-\epsilon_{l}\epsilon_i\sin(\theta)&\cos(\theta)&0\\0&0&1\end{pmatrix}\begin{pmatrix}S_{li}^x\\S_{li}^y\\S_{li}^z\end{pmatrix},
\end{eqnarray}
to remove the twists in each layer. Then we obtain the twists on the rung links as follows
\begin{eqnarray}
  \label{eq:tsm}
  H&=& J\sum_{\langle ij\rangle l}\tilde{\mathbf{S}}_{li}\cdot\tilde{\mathbf{S}}_{lj}+J_{\bot}\sum_i[\cos(2\theta)\tilde{\mathbf{S}}_{1i}\cdot\tilde{\mathbf{S}}_{2i}\nonumber\\
    &&+2\sin^2(\theta)\tilde{S}_{1i}^z\tilde{S}_{2i}^z+\epsilon_i\sin(2\theta)(\tilde{\mathbf{S}}_{1i}\times\tilde{\mathbf{S}}_{2i})\cdot\hat{z}].
\end{eqnarray} 
When $J_{\bot}=0$, the model (\ref{eq:tsm}) describes two decoupled SU(2)-invariant pseudospin-1/2 models as that in the single layer Sr$_2$IrO$_4$ which has the canted antiferromagnetic structure and shows ferromagnetism with large ferromagnetic moment\cite{Cao1998,Wang2011}. If we take the $J_{\bot}$ as the perturbation, we may anticipate the in-plane canted antiferromagnetism in the bilayer Sr$_3$Ir$_2$O$_7$. The presence of the interlayer coupling prevents the full elimination of the twists in the bilayer twisted pseudospin model and brings the intrinsic anisotropy in Sr$_3$Ir$_2$O$_7$ which selects  the easy $c$-axis AF magnetic structure different from that of Sr$_2$IrO$_4$. When $J_{\bot}\neq0$, however, the pseudospin alignment along $c$-axis gains more energy from the interlayer couplings than the in-plane pseudospin alignment. The $c$-axis is the easy axis for the magnetic moment alignment. Here we give alternative explanation of the  easy $c$-axis different from that in Ref. \cite{Kim2012a}. Due to intrinsic anisotropy of the bilayer twists in Sr$_3$Ir$_2$O$_7$, the low temperature AF order breaks the discrete Ising symmetry (not SU(2)) and the transverse spin wave excitations are gapped. The interlayer coupling $J_\bot$ also introduces the quantum fluctuations and reduces the magnetic moment.  We will use the bond-operator mean field method to quantitatively describe these properties.

\textit{Bond-operator mean field method.}
It has been proved that the bond-operator representation is a very convenient method to treat  the bilayer spin system\cite{Sachdev1990,Sachdev2004}. For the rungs along $c$-axis,  we introduce the bond operators\cite{Sachdev1990} (singlet boson $s$ and triplet boson $t_{x,y,z}$), $|s\rangle\equiv s^\dag|0\rangle= \frac{1}{\sqrt{2}}(|\uparrow\downarrow\rangle-|\downarrow\uparrow\rangle)$, $|t_x\rangle\equiv t_x^\dag|0\rangle=\frac{-1}{\sqrt{2}}(\uparrow\uparrow\rangle-|\downarrow\downarrow\rangle)$, $|t_y\rangle\equiv t_y^\dag|0\rangle=\frac{i}{\sqrt{2}}(|\uparrow\uparrow\rangle+|\downarrow\downarrow\rangle)$ and $|t_z\rangle\equiv t_z^\dag|0\rangle= \frac{1}{\sqrt{2}}(|\uparrow\downarrow\rangle+|\downarrow\uparrow\rangle)$
to describe the four spin states ($|\uparrow\uparrow\rangle$, $|\uparrow\downarrow\rangle$ $|\downarrow\uparrow$, and $|\downarrow\downarrow\rangle$). On every rung, we have the constraint $s^\dag s+\sum_{\alpha}t_{\alpha}^\dag t_{\alpha}=1$ ($\alpha=x,y,z$).  

\begin{figure}[b]
  \begin{center}
    \includegraphics[width=0.5\columnwidth]{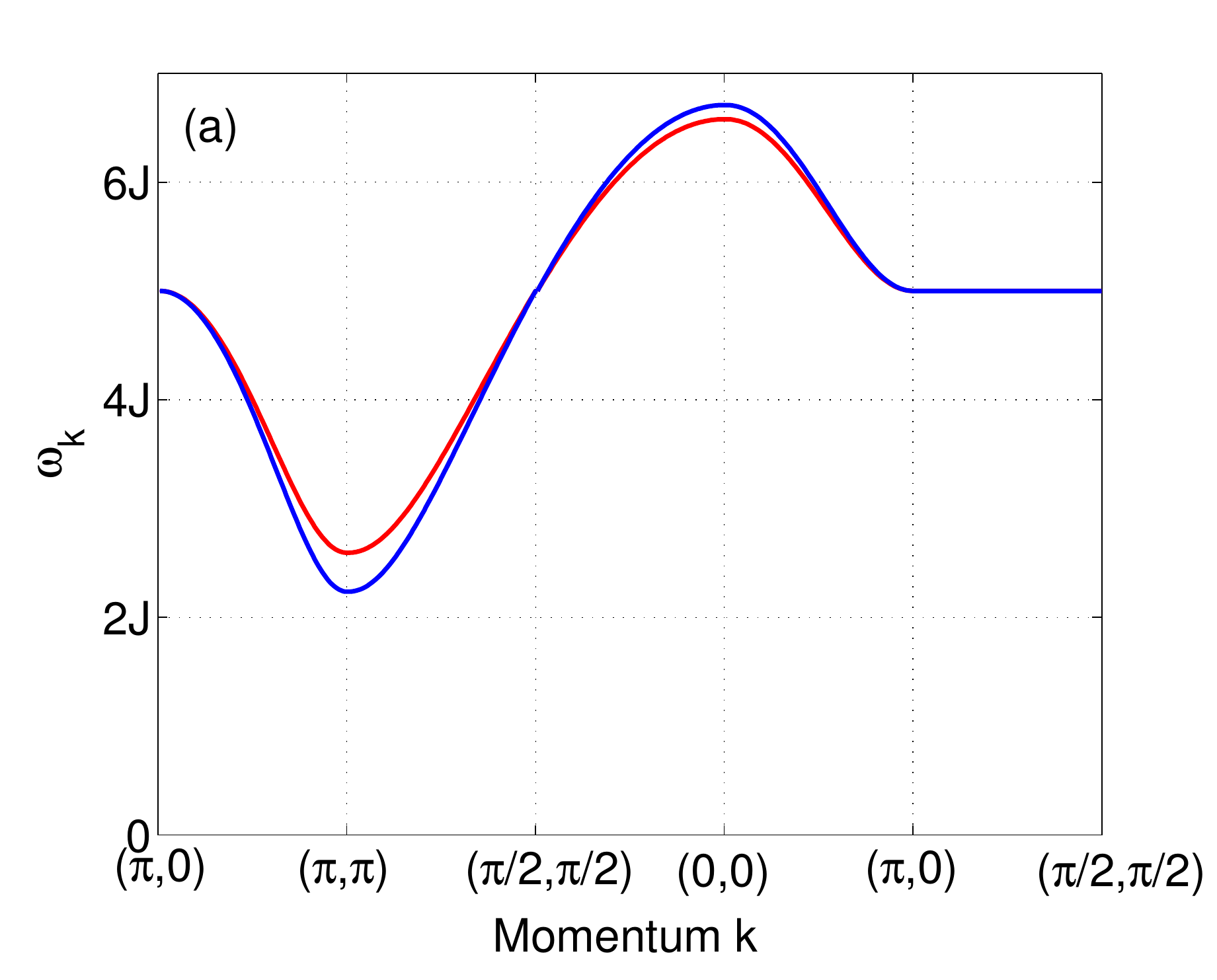}\includegraphics[width=0.5\columnwidth]{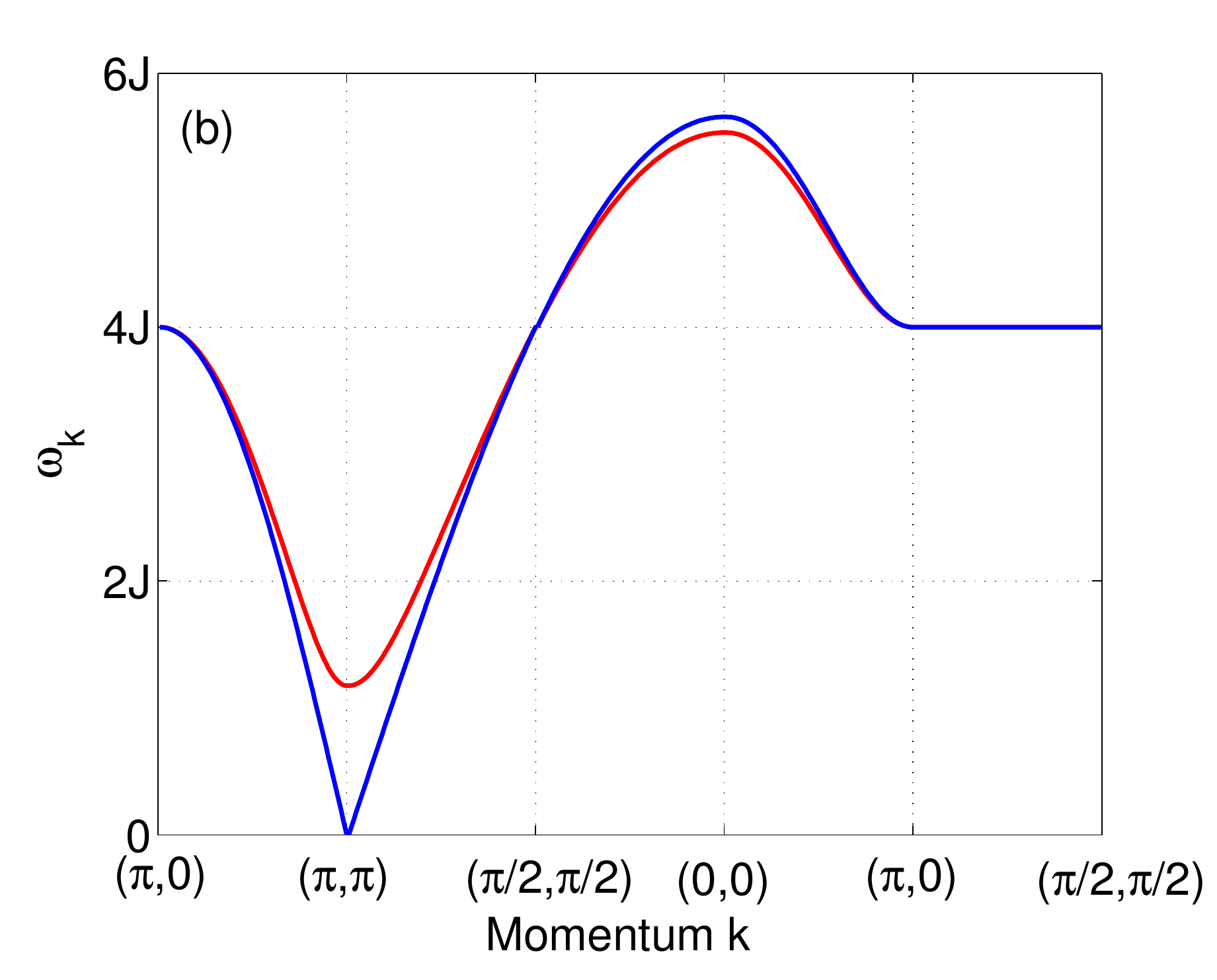}\\
\includegraphics[width=0.5\columnwidth]{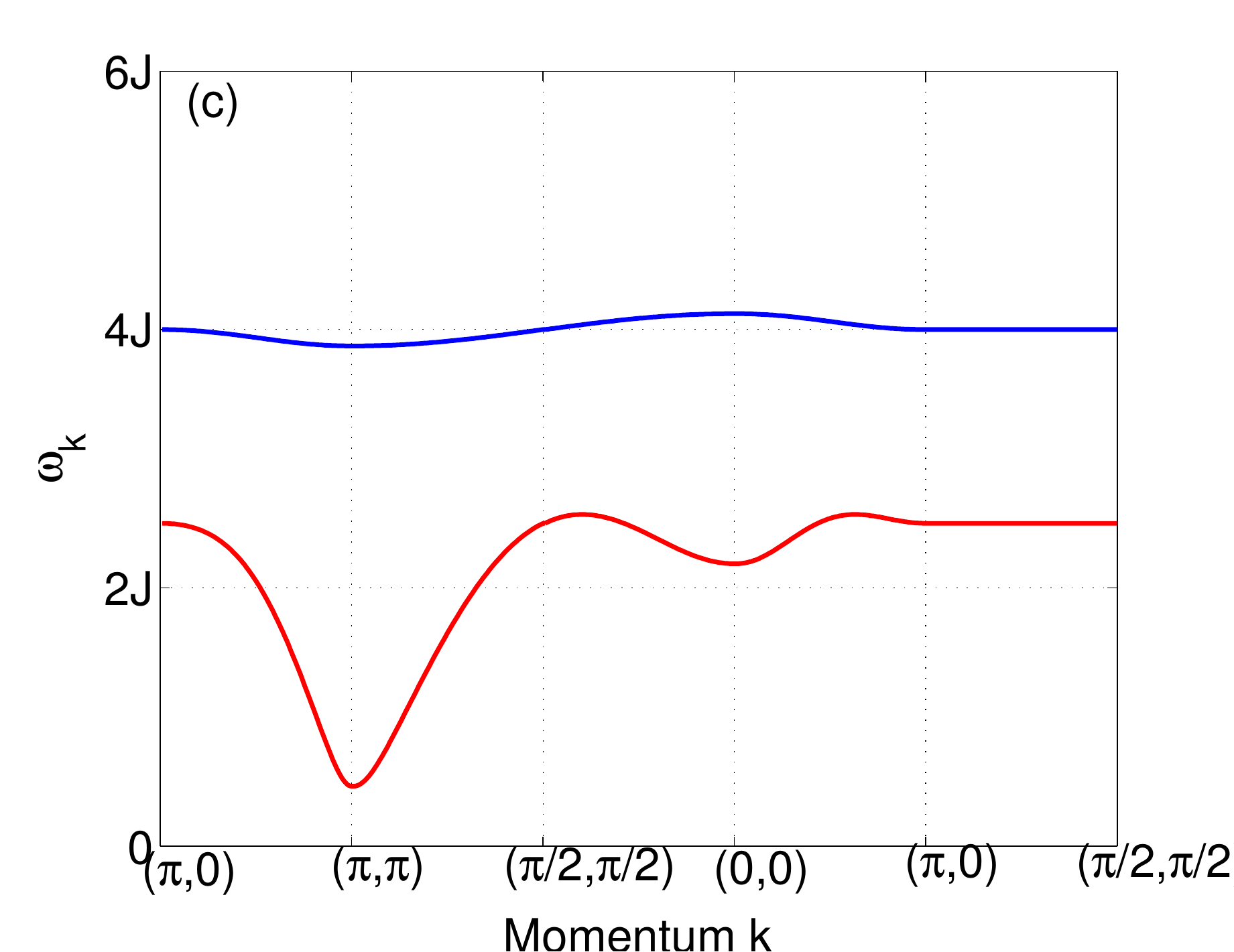}\includegraphics[width=0.5\columnwidth]{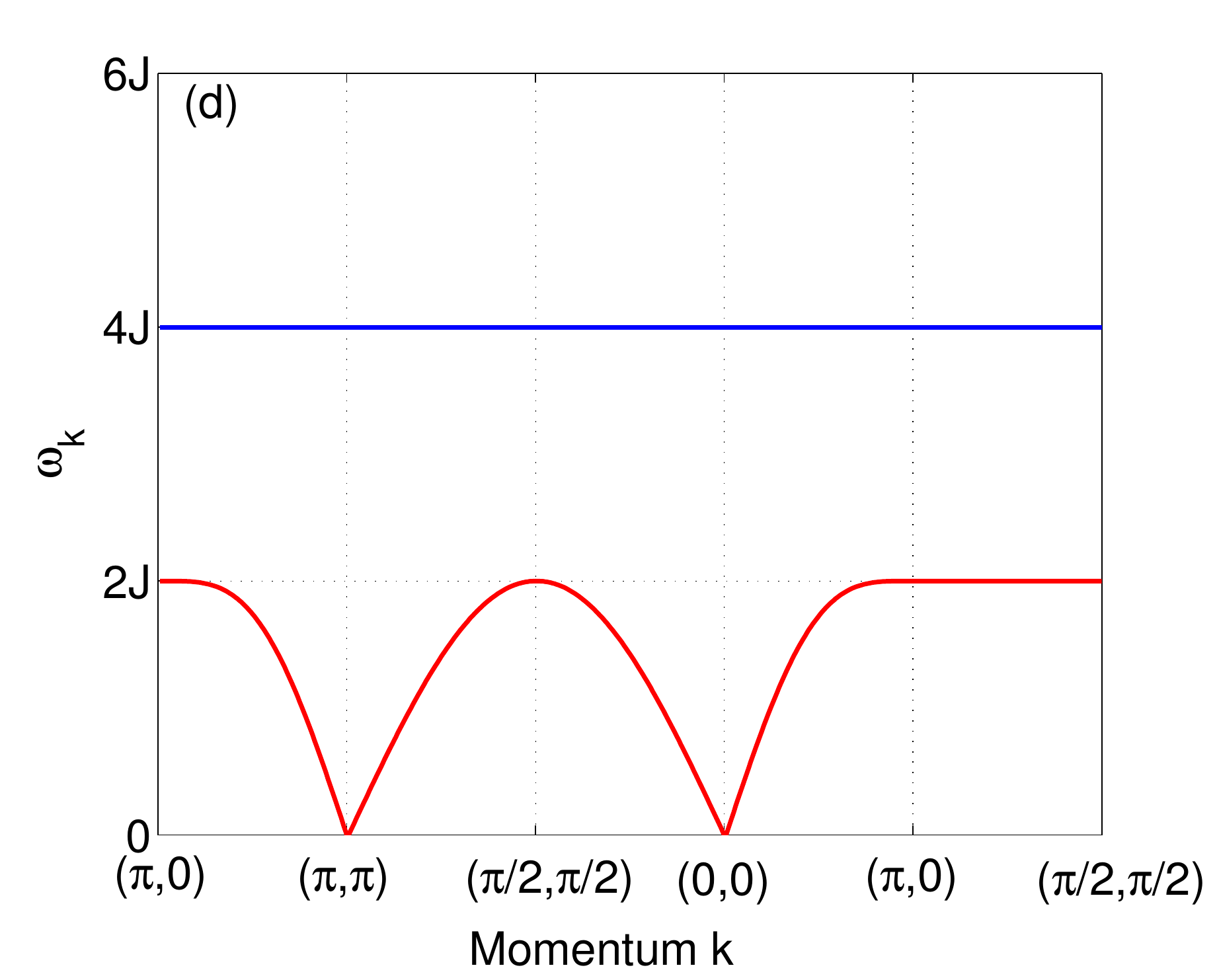}
  \end{center}
  \caption{Mean field dispersions of the bilayer Heisenberg spin waves for different values of $J_{\bot}$: (a) $J_{\bot}=5J$, (b) $J_{\bot}=4J$, (c) $J_{\bot}=J$, (d) $J_{\bot}=0$. The red (blue) line is for $t_{x,y}$ ($t_z$) bosons.} 
  \label{fig:sw}
\end{figure}
To simplify the calculations, we implement the mean field calculation at zero temperature and vary the interlayer couplings to tune the fluctuations.  The singlet boson $s$ condenses in all phase and the condensation of the triplet bosons describes the magnetic order. We start from the disorder state (large $J_{\bot}$ limit) and then decrease the fluctuations to see the magnetic instability.  In the disorder state, we implement the approximation for the ground state, $|G\rangle=\prod_i s_i^\dag|0\rangle$, and obtain the mean field Hamiltonian as 
\begin{eqnarray}
  H_{\text{MF}}&=&\sum_{i\alpha}J_{\bot}t_{i\alpha}^\dag t_{i\alpha}+\frac{J}{2}\sum_{\langle ij\rangle}({t}_{iz}^\dag {t}_{jz}+{t}_{iz}^\dag {t}_{jz}^\dag+\text{h.c.})\nonumber\\
  &+&\frac{J}{2}\sum_{\langle ij\rangle}(\cos(2\theta)(t_{i\tau}^\dag t_{j\tau} -t_{i\tau}^\dag t_{j\tau}^\dag+\text{h.c.}),
\end{eqnarray}
with $\alpha=x,y,z$ and $\tau=x,y$.  The mean field Hamiltonian can be diagonalized as $H_{\text{MF}}=\sum_{\mathbf{k}\alpha}\omega_{\mathbf{k}\alpha}\xi_{\mathbf{k}\alpha}^\dag \xi_{\mathbf{k}\alpha}$ with the spin wave dispersions $\omega_{\mathbf{k}\alpha}=\sqrt{\Lambda_{\mathbf{k}\alpha}^2-\Gamma_{\mathbf{k}\alpha}^2}$. Here we have $\Lambda_{\mathbf{k}\tau}= J_{\bot}+\frac{J\cos(2\theta)}{2}\gamma_\mathbf{k}$, $\Gamma_{\mathbf{k}\tau}= \frac{J\cos(2\theta)}{2}\gamma_\mathbf{k}$, $\Lambda_{\mathbf{k}z}= J_\bot+\frac{J}{2}\gamma_\mathbf{k}$, and $\Gamma_{\mathbf{k}z}=\frac{J}{2}\gamma_\mathbf{k}$
with $\gamma_{\mathbf{k}}=2(\cos(k_x)+\cos(k_y))$.  
Due to the intrinsic anisotropy of the twists, $t_z$ has the smaller gap than $t_{x,y}$ at AF wave vector in the disordered state when $J_{\bot}=5J$ as shown in Fig. \ref{fig:sw} (a). Decreasing the fluctuations ($J_{\bot}$),  $t_z$ becomes gapless when $J_{\bot}=4J$ as shown in Fig. \ref{fig:sw} (b). 

With further decreasing $J_{\bot}$, $t_z$ condenses at AF wave vector and the system develops the easy $c$-axis collinear AF magnetic order. The AF order state is described  as $|G\rangle=\prod_i\tilde{s}_i^\dag|0\rangle$ with the new basis states, $\tilde{s}_i^\dag= \cos(\chi)s_i^\dag-\epsilon_i\sin(\chi)t_{iz}^\dag $ and $\tilde{t}_{iz}^\dag= \epsilon_i\sin(\chi)s_i^\dag+\cos(\chi)t_{iz}^\dag$\cite{Sommer2001}. Now $\tilde{t}_z$ is the longitudinal spin wave excitation while $t_{x}$ and $t_y$ are the transverse ones.  The mean field Hamiltonian is now
\begin{eqnarray}
  H_{\text{MF}}&=& J\sum_i2\epsilon_i\sin(2\chi)(\cos(2\chi)-\frac{J_{\bot}}{4J})(\tilde{t}_{iz}^\dag+\tilde{t}_{iz})\nonumber\\
  &+&4J\sum_i[\frac{J_{\bot}}{4J}\cos(2\chi)+\sin^2(2\chi)]\tilde{t}_{iz}^\dag \tilde{t}_{iz}\nonumber\\
  &+&\frac{J}{2}\sum_{\langle ij\rangle}\cos^2(2\chi)(\tilde{t}_{iz}^\dag \tilde{t}_{jz}+\tilde{t}_{iz}^\dag \tilde{t}_{jz}^\dag+\text{h.c.})\nonumber\\
  &+&2J\sum_{i\tau}[\frac{J_{\bot}}{2J}\cos^2(\chi)+\sin^2(2\chi)](t_{i\tau}^\dag t_{i\tau}+\text{h.c.})\nonumber\\
  &+&\frac{J}{2}\sum_{\langle ij\rangle \tau}[(\cos(2\theta)\cos(2\chi)t_{i\tau}^\dag t_{j\tau}\nonumber\\
    &-&(\cos(2\theta)+i\sin(2\theta)\sin(2\chi)t_{i\tau}^\dag t_{j\tau}^\dag))+\text{h.c.}].
\end{eqnarray}
with $\tau=x,y$. Since $|G\rangle$ is the ground state, the coefficient of the longitudinal spin excitation vanishes to give a self-consistent description, $\sin(2\chi)(\cos(2\chi)-\frac{J_{\bot}}{4J})=0$.
When $J_{\bot}<4J$, the system is in the magnetic ordered state and $\cos(2\chi)=\frac{J_\bot}{4J}$.  The spin wave excitations are $\omega_{\mathbf{k}\alpha}=\sqrt{\Lambda_{\mathbf{k}\alpha}^2-\Gamma_{\mathbf{k}}^2}$ with $\Lambda_{\mathbf{k}\tau}= (2J+\frac{J_\bot}{2})+\frac{J_\bot\cos(2\theta)}{8}\gamma_\mathbf{k}$, $\Gamma_{\mathbf{k}\tau}= \frac{J}{2}\sqrt{1-\frac{J_\bot^2\sin(2\theta)^2}{16J^2}}\gamma_\mathbf{k}$, $\Lambda_{\mathbf{k}z}= 4J+\frac{J_{\bot}^2}{32J}\gamma_\mathbf{k}$ and $\Gamma_{\mathbf{k}z}=\frac{J_\bot^2}{32J}\gamma_\mathbf{k}$. When $0<J_\bot<4J$, the transverse spin wave excitations have the gap
\begin{eqnarray}
  \Delta_t=\sqrt{(4J+J_\bot)J_\bot}\sin(\theta).
\end{eqnarray}
For Sr$_3$Ir$_2$O$_7$, $\theta=12$\textdegree ~and $J_{\bot}\simeq J\simeq 80$ meV. As shown in Fig. \ref{fig:sw} (b), the spin gap  is $\Delta_t\simeq 40$ meV. In the inelastic x-ray scattering, a weak and broad feature above the transverse band is observed and interpreted as the two-magnon scattering\cite{Kim2012b}. Compared with the Fig. \ref{fig:sw} (c), the weak and broad feature may originate form the weak dispersive longitudinal spin wave excitation. The AF moment is $m_{\text{AF}}(J_\bot=J)/m_{\text{AF}}(0)=0.968$. When $J_{\bot}=0$, the bond-operator mean field method gives a self-consistent description of the decoupled bilayer model. The transverse spin wave excitations are gapless at the AF wave vector and the longitudinal mode is non-dispersive, as shown in Fig. \ref{fig:sw} (d).

The quantum fluctuations due to the hardcore condition for the $t_\alpha$ bosonic fields are ignored in the mean field treatment\cite{Kotov1998}. The mean field critical interlayer coupling $J_{\bot}^c=4J$ is always larger than the exact one.  The spin gap and the moment reduction are significantly underestimated comparing to the experiments\cite{Kim2012b,Fujiyama2012}. Our main purpose in this paper is to give a qualitative description and the exact treatment  is left for further investigations.

\textit{Summary.} In this paper, we use the bilayer twisted Hubbard model as the zeroth order approximation of the low energy electronic structure of Sr$_3$Ir$_2$O$_7$. After simple treatments, this model can give the good understanding of following properties: (i) At low temperature, Sr$_3$Ir$_2$O$_7$ has the easy $c$-axis collinear long-range AF order.  (ii) The transverse spin wave excitations are gapped and the longitudinal one is weakly dispersive. (iii) The magnetic moment is reduced by the interlayer couplings.  The bilayer twisted Hubbard model gives a good description of low energy electronic properties in Sr$_2$Ir$_3$O$_7$.

J. W. Mei thanks T. M. Rice for useful and encouraging discussions during the project. He also thanks S. Fujiyama for the discussion. The work is supported by Swiss National Fonds.
\bibliography{SrIrO}
\end{document}